\documentclass[conference]{IEEEtran}
\usepackage[utf8]{inputenc}

\DeclareUnicodeCharacter{0307}{\.}
\IEEEoverridecommandlockouts
\usepackage{cite}
\usepackage{amsmath,amssymb,amsfonts}
\usepackage{algorithmic}
\usepackage{graphicx}
\usepackage{textcomp}
\usepackage{xcolor}
\usepackage{epsfig}
\usepackage{epstopdf}
\usepackage{balance}
\usepackage[printonlyused]{acronym}
\usepackage{booktabs}
\usepackage{tabularx}
\usepackage{dirtytalk}
\usepackage{url}
\acrodef{thz}[THz]{Terahertz}
\acrodef{uav}[UAVs]{unmanned aerial vehicles}
\acrodef{snr}[SNR]{signal-to-noise ratio}
\acrodef{los}[LoS]{line-of-sight}
\acrodef{hpbw}[HPBW]{half power beam width}
\newcommand{\FGR}[1]{Fig.~\ref{#1}}

\newcommand{\SEC}[1]{Section~\ref{#1}}
\newcommand{\TAB}[1]{Table~\ref{#1}}
\def\BibTeX{{\rm B\kern-.05em{\sc i\kern-.025em b}\kern-.08em
    T\kern-.1667em\lower.7ex\hbox{E}\kern-.125emX}}
\begin{document}

\title{
% Measurement Based Analysis of the Misalignment on THz Frequencies başlkk güncellenecek
% Investigating the Effect of Misalignment for UAV-assisted THz Communication via Empirical Measurements
Experimental Assessment of Misalignment Effects in Terahertz Communications
}
\author{\IEEEauthorblockN{Hasan Nayir\IEEEauthorrefmark{1}, Erhan Karakoca\IEEEauthorrefmark{1}\IEEEauthorrefmark{2}, G\"{u}ne\c{s} Karabulut Kurt\IEEEauthorrefmark{3}, Ali G\"{o}r\c{c}in\IEEEauthorrefmark{1}\IEEEauthorrefmark{2}
}
\IEEEauthorblockA{\IEEEauthorrefmark{1} Department of Electronics and Communication Engineering, {\.{I}}stanbul Technical University, {\.{I}}stanbul, Turkey}
\IEEEauthorblockA{\IEEEauthorrefmark{2}Communications and Signal Processing Research (HİSAR) Laboratory, T{\"{U}}B{\.{I}}TAK B{\.{I}}LGEM, Kocaeli, Turkey}
\IEEEauthorblockA{\IEEEauthorrefmark{3}Poly-Grames Research Center, Department of Electrical Engineering, Polytechnique Montréal, Montréal, QC, Canada}
\\ 
Emails: \texttt{\{nayir20, karakoca19\}@itu.edu.tr,gunes.kurt@polymtl.ca,}\\
\texttt{{ali.gorcin@tubitak.gov.tr}}}
\maketitle
\begin{abstract}
% Terahertz (THz) frequencies are important for next generation wireless systems due to the advantages in terms of large available bandwidths. On the other hand, the limited range due to high attenuation in these frequencies can be overcome via densely installed heterogeneous networks also utilizing UAVs in a three-dimensional hyperspace. Yet, THz communications rely on precise beam alignment, if not handled properly results in low signal strength at the receiver which impacts THz signals more than conventional ones. This work focuses on the importance of precise alignment in THz communication systems and the significant effect of proper alignment is validated through comprehensive measurements conducted through a state-of-the-art measurement setup, which enables accurate data collection between 240 GHz to 300 GHz at varying angles and distances in an anechoic chamber eliminating reflections. By analyzing the channel frequency and impulse responses of these extensive and particular measurements, this study provides the first quantifiable results in terms of measuring the effects of beam misalignment in THz frequencies.

Terahertz (THz) frequencies play a crucial role in the advancement of next-generation wireless systems, primarily owing to their substantial available bandwidths. The inherent limitation of limited range, attributed to high attenuation in these frequencies, can be effectively addressed by implementing densely deployed heterogeneous networks, complemented by Unmanned Aerial Vehicles (UAVs) within a three-dimensional hyperspace.
Yet, the success of THz communications relies on the precise alignment of beams. Inadequate handling of beam alignment can lead to diminished signal strength at the receiver, significantly affecting THz signals more than their conventional counterparts. This research underscores the paramount importance of meticulous alignment in THz communication systems. The profound impact of proper alignment is substantiated through comprehensive measurements conducted using a state-of-the-art measurement setup, facilitating accurate data collection across the 240 GHz to 300 GHz spectrum. These measurements encompass varying angles and distances within an anechoic chamber to eliminate reflections.
Through a meticulous analysis of the channel frequency and impulse responses derived from these extensive measurements, this study pioneers quantifiable results, providing an assessment of the effects of beam misalignment in THz frequencies.

\end{abstract}

\section{Introduction}
While the need for high data rates is still on the agenda, the 6G vision has highlighted various key value indicators (KVIs) such as global coverage, service availability, sustainability, and reliability. When we set out with the \say{connection in anywhere, anytime, any device} motto, there is no doubt that aerial systems will be the most prominent candidate to bring access to urban, semi-urban, and remote rural areas. Adding aerial base stations for improving the quality of service (QoS) and boosting the coverage, reliability and capacity of wireless networks has been suggested in the academy for a while~\cite{7417609,sobouti2023managing,giordani2020toward}. In addition to bringing fast deployment features to non-terrestrial networks (NTNs), \ac{uav} also acts as a bridge between terrestrial networks (TNs) and other non-terrestrial network elements such as satellites and high altitude platforms (HAPs). 

\ac{thz} wireless systems are expected to be a vital enabler for 6G in tandem with TNs and NTNs because of their large contiguous bandwidth~\cite{boulogeorgos2018terahertz} which allows them to keep pace with the surge in wireless data volume and the increasing amount of traffic as new nodes are added to the network.
Likewise, considering the fact that \ac{thz} bands are not allocated yet for specific active services around the globe, there is an enormous potential to meet the need for the desired communication traffic. Hence the orchestration of NTNs and TNs with the THz communication is apparent towards 6G\cite{9269928}.

Along with their benefits, \ac{thz} frequencies also come with high attenuation due to molecular absorption and spreading loss \cite{jornet2011channel}, which limits the communication range significantly. 
% Thus, densely deployable innovative THz communication systems are required to cope with this issue. 
Addressing these limitations demands innovative THz communication systems. This is where UAVs come into play as a remedy especially to provide instant high-capacity communication links in crowded environments or to support high-capacity data traffic between different TN and NTN nodes~\cite{azari2022thz}.
% and Unmanned Aerial Vehicles (\ac{uav}) emerge as a transformative solution. 
% Along with their, cost-effectiveness, 3D deployment capabilities ensuring Line-of-Sight (\ac{los}) conditions, and flexibility for network nodes, \ac{uav} stand as instrumental allies in advancing \ac{thz} communications, unlocking new possibilities and services~\cite{azari2022evolution}.
Moreover, along with their cost-effectiveness and instant 3D deployment capabilities which allow maintaining \ac{los} condition for the communication links, \ac{uav} also provides flexibility to the network nodes. \ac{uav} and \ac{thz} are strong collaborators by their nature and this mutually constructive relationship, in turn, can unlock new opportunities and innovative services\cite{azari2022evolution}. 
Thus, \ac{uav} are expected to pave radically the way for assisting \ac{thz} communications.

While \ac{thz}-integrated \ac{uav} presents promising prospects they also bring new challenges to the field.
Although the utilization of directional beamforming unified with directional antennas can provide higher antenna gains to reduce the high transmission loss in the \ac{thz} frequency range, these systems are prone to pointing errors due to small beamwidths. 
Moreover, wind or sudden complex movements can cause uncontrollable tilts or rotations in UAV operations, leading to beam misalignment and an inevitable decrease in \ac{snr}.
Accordingly, UAV-assisted THz communication requires accurate beam alignment mechanisms and algorithms.
As we move towards the development of 6G networks and the realization of \ac{thz} communication systems, it is crucial to conduct a comprehensive investigation of misalignment scenarios. This involves analyzing and modeling the potential effects of the misalignment on specific applications.

\subsection{Related Works}
The antenna misalignment effects on \ac{thz} communication systems have been investigated across various environments and frequency ranges, but mostly by simulations~\cite{priebe2012impact, priebe2012affection, papasotiriou2020performance, badarneh2022performance}. In particular, \cite{priebe2012impact} and \cite{priebe2012affection} examined the effects of antenna misalignment at 300 GHz in a simulated office setup by considering practical propagation conditions. In \cite{papasotiriou2020performance},  the performance of a multicarrier \ac{thz} wireless system is evaluated under the fading condition caused by misalignment. The effects of pointing error impairments under random fog conditions are examined in~\cite{badarneh2022performance}.

On the other hand, measurement based impact of misalignment has been analysed in \cite{ekti2017statistical, sheikh2021horn}.
The authors in \cite{ekti2017statistical} carried out measurements to analyze the impact of the distance and single degree of misalignment on the path loss in the THz communication system and several important statistical parameters for line–of–sight (LOS) channels are measured.
The performance of the experimental \ac{thz} communication systems has been examined in case of antenna misalignment at 100, 300, 400 and 500 GHz by utilizing proper horn antennas in the \cite{sheikh2021horn}. 
A significant decrease trend was indicated in the received power with misalignment due to the divergence of the beams particularly with an increase in separation distance.
Also, the new modelling technique for measurement-based THz channels is proposed in \cite{9852427} and regarding this model capacity analysis is given in \cite{10437716}. 
Most importantly, the authors in \cite{guan2019effects} have designed a drone-based measurement setup to investigate the effects of mobility uncertainties on mmWave/\ac{thz}-band communications between flying drones. The authors showed that the mobility of the UAVs when they are in movement causes significant performance degradation and link outages while propeller rotation and engine operations of the UAVs cause far less performance degradation.
% In \cite{dabiri2022pointing}, the authors concentrate on developing an analytical framework to accurately assess and measure the pointing error in millimeter wave (mmWave) and terahertz (THz) communication links. They consider misalignment as a source of randomness that has an impact on antenna gains. 

% Also, some researchers have taken a more in-depth look at the misalignment problem in free space optical communication links and have come up with a more complete model that takes into consideration not just misalignment but also other factors, such as turbulence in the atmosphere and how small movements (displacement jitter) or receiver detector size \cite{farid2007outage}. In addition, this model was adopted by \cite{boulogeorgos2019analytical, kokkoniemi2020impact}.

% Furthermore, measurement, modeling, and analysis for terahertz wireless channels are presented extensively in \cite{han2022terahertz}.
% [buraya bişey lazım!!!!] 

\subsection{Contributions}
In order to fully maximize the potential of \ac{thz} communication systems, a deep understanding of their performance in practical conditions is required. As the realization of the UAV-assisted \ac{thz} communication becomes increasingly prevalent, the effect of misalignment becomes a more prominent concern. 
% While prior studies have explored the impact of distance and antenna misalignment to some extent on \ac{thz} communication more comprehensive approaches ought to address this issue. 
Addressing this issue necessitates a more comprehensive approach, as prior studies have only partially explored the impact of distance and antenna misalignment on THz communication.
To achieve this, it is essential to gather application-specific measurements and conduct an in-depth analysis of their impact on channel frequency and impulse response. 
In the light of these motivations, our contributions are listed as follows: 
%Pointing out that, we developed a state-of-the-art measurement setup to collect precise measurements for closer inspection of the effect of the misalignment with the distance on UAV-assisted THz communication systems. 
\begin{itemize}
    \item Undertaking precise controlled THz misalignment experiments is a formidable task that requires extensive expertise and experience. To unravel the intricacies involved, a groundbreaking measurement system has been devised to capture precise measurements under varying misalignment scenarios and at different distances. Moreover, this study strives to illuminate future research endeavours by providing a comprehensive explication of the measurement campaign and sharing invaluable insights derived from a multidisciplinary approach.
    % \item Conducting precise THz misalignment experiments is an arduous undertaking that demands considerable expertise and experience. To understand different peculiarities a novel measurement system has been developed to take precise measurements from different misalignments at various distances. Moreover, this study endeavours to illuminate forthcoming research endeavours by elucidating the measurement setup and sharing the invaluable insights gleaned from a comprehensive, multidisciplinary approach.
    \item Also, recognizing the significance of large bandwidths in practical THz communication systems, this study stands apart from the majority of existing literature on THz measurements. Rather than focusing solely on a specific frequency range, measurements were conducted using a single scan method, encompassing the 60 GHz band and spanning from 240 GHz to 300 GHz.
    \item The measurements were analyzed in terms of the channel frequency and impulse responses to gain insights into the joint effect of the distance and misalignment. 
    % \item Performing controlled THz misalignment experiments is a very grueling and experience-requiring process. This study also aims to shed light on future studies by explaining the experiment setup and conveying insights obtained through the long and multidisciplinary approach. 
\end{itemize}

% Maximizing the potential of \ac{thz} communication systems requires a deep understanding of their performance in practical scenarios. As the realization of the \ac{uav}-assisted \ac{thz} communication becomes increasingly prevalent, the effect of misalignment becomes prominent. Thus, gathering application-specific measurements and analyzing their impact on channel frequency and impulse response is crucial to address this issue.
% While the majority of the works have been done analytically to investigate the effects of the different peculiarities, this study focuses on the impact of distance and antenna misalignments on \ac{thz} communication by conducting measurements. Even though there is several works have been done to investigate this effect, there are no comprehensive measurement based works. For this purpose, an application-specific measurement setup is used, which allows the collection of measurements with precise adjustments in different distance and misalignment settings. 
% Also, considering the high bandwidth opportunity of the \ac{thz} frequencies, the measurements are taken in the 60 GHz band between 240 GHz and 300 GHz by using the single sweep method. 
% The collected data is then analyzed by the channel frequency and impulse responses to gain insights about the joint effect of the distance and misalignment. 

% \begin{figure}
%     \centering
%     \includegraphics[width=\linewidth]{figures/beam_drone.png}
%     \caption{ UAV-assisted THz wireless communication link with prospective misalignment contributors.}
%     \label{fig:my_label}
% \end{figure}

\subsection{Organization}
% The remainder of this paper is organized as follows: \SEC{system_model} firstly introduces the signal model and then explains the measurement campaign. Measurement results are given in \SEC{measurement_results}. Finally, the conclusion and future direction are presented in \SEC{conclusion}.  
The remainder of this paper is organized as follows: in \SEC{system_model}, the signal model is introduced, providing a foundational understanding of the system. Subsequently, the measurement campaign is explained, outlining the approach and methodology employed. The obtained measurement results are presented in \SEC{measurement_results}, showcasing valuable findings and insights. Finally, the conclusion and future directions are discussed in \SEC{conclusion}, summarizing the key outcomes and presenting potential avenues for further research.

\section{Signal Model and System Overview}\label{system_model}
In this study, the effect of misalignment for \ac{thz} channels is investigated in the anechoic chamber. 
% For the analysis part, the traditional linear, time-invariant channel model approach is adopted to be simplified. 
The traditional linear, time-invariant channel model approach is used to reduce complexity in the analysis.

\subsection{Signal Model}
The received signal at the passband can be expressed as
\begin{equation}
r(t)=\operatorname{Re}\left\{\left(x_I(t)+j x_Q(t)\right) e^{j 2 \pi f_c t}\right\},
\end{equation}
where $f_{c}$ denotes the carrier frequency. Also, $x_I(t)$ and $x_Q(t)$ are the in-phase and quadrature components of the received signal, respectively. The received signal can be modeled as a superposition of multipath signals with different delays and complex gains. So, the  channel at the baseband can be represented as 
\begin{equation}\label{channel_baseband}
h(t)=\sum_{l=0}^{L-1} \alpha_l e^{-j 2 \pi f_c t_l} \delta\left(t-t_l\right),
\end{equation}
where $L$, $\alpha_{l}$, and $t_{l}$ represent the number of multipath components, channel complex gain, and delay for the $l$-th path, respectively. As we have mentioned before, measurements have been carried out in a fully isolated anechoic chamber, so we can assume there is only \ac{los} signal transmission. Thus, \ac{los} channels can be derived for $L=1$ (\ref{channel_baseband}) as 
\begin{equation}
h(t)=a_f e^{j \theta} \delta\left(t-t_0\right),
\end{equation}
where $a_f$, $\theta$, and $t_{0}=d/c$ denote the \ac{los} path complex gain,  phase of the signal, and propagation delay, respectively. Also, $d$ is the distance between the transmitter and receiver, and $c$ is the speed of the light. It is crucial to acknowledge that when using directional antennas, which is a common practice in THz communication, the effects of antenna misalignment, frequency-dependent loss, and frequency dispersion index can all be accounted for by the term $a_f$.

In existing literature, the stochastic characterization of multipath components in a static environment is commonly regarded as a combination of specular and diffused components, forming a superposition \cite{yarkan2008identification}.
\begin{subequations}
\begin{equation}
\begin{aligned}
m_l & =a_l e^{-i 2 \pi f_{c t}} \\
& =s_l+d_l
\end{aligned}
\end{equation}

\begin{equation}
s_l=\sigma_{s_l} e^{\left(j 2 \pi f_0 \cos \left(\theta_l\right)+\phi_l\right)}
\end{equation}

\begin{equation}
d_l=\sigma_{d_l} \frac{1}{\sqrt{M_l}} \sum_{m=1}^{M_l} b_m e^{\left(j 2 \pi f c \cos \left(\theta_m\right)+\phi_m\right)}
\end{equation}
\end{subequations}
where the term $\sigma_{s_l}$ represents the magnitude of the specular component, while $\theta_l$ denotes the angle of arrival (AoA) and $\phi_l$ represents the phase of the specular component. Similarly, $\sigma_{d_l}$ corresponds to the magnitude of the diffused component, $M_{l}$ signifies the number of diffused waves, $b_{m}$ represents the amplitude of the incoming waves, $\theta_{m}$ denotes the AoA, and $\phi_{m}$ is the phase of the incoming waves forming the diffused component, respectively. It is commonly assumed, without loss of generality, that both $\sigma_{s_l}$ and $\sigma_{d_l}$ can be considered as unity under ideal conditions.

The \ac{los} scenario stands out as a special case in wireless propagation, exhibiting inherent characteristics in both large-scale and small-scale fading mechanisms. To guarantee \ac{los} transmission, it is crucial to implement a fully isolated measurement setup within an anechoic chamber, incorporating absorbers. This setup effectively limits the losses introduced by the propagation channel to factors such as distance-dependent path loss, potential antenna misalignments, equipment imperfections, and non-ideal behaviors that may arise when operating in proximity to or above ideal conditions.

In this model, the losses are contingent upon both the distance and misalignment between the transmitter and receiver. When taking into account the distance and angular losses associated with a directional antenna having a maximum gain direction angle $\varphi$, the loss can be expressed in decibels dB as follows:

\begin{equation}
P L=P L_0+10 n \log _{10}(d)+\mathbb{G}\left(\phi-\varphi\right),
\end{equation}
$d$ is the distance between the extender modules. $\mathbb{G}$ represents the normalized angular gain pattern of an antenna and $\phi$ is the angle between the extender modules.

The angular gain function in linear scale can be approximated as \cite{zang2010bayesian}
\begin{equation}
G(\theta)=\left|\frac{\sin (\omega \theta)}{\omega \theta}\right|, \quad|\theta| \leq \pi,
\end{equation}
where $\omega$ is a parameter linked to both the maximum gain direction angle and the beamwidth of the directional antenna. The antenna beamwidth can be defined as twice the angular value at which the measured power in the maximum gain direction decreases by half.

\begin{figure*}[!t]
   \centering    \includegraphics[width=0.8\linewidth]{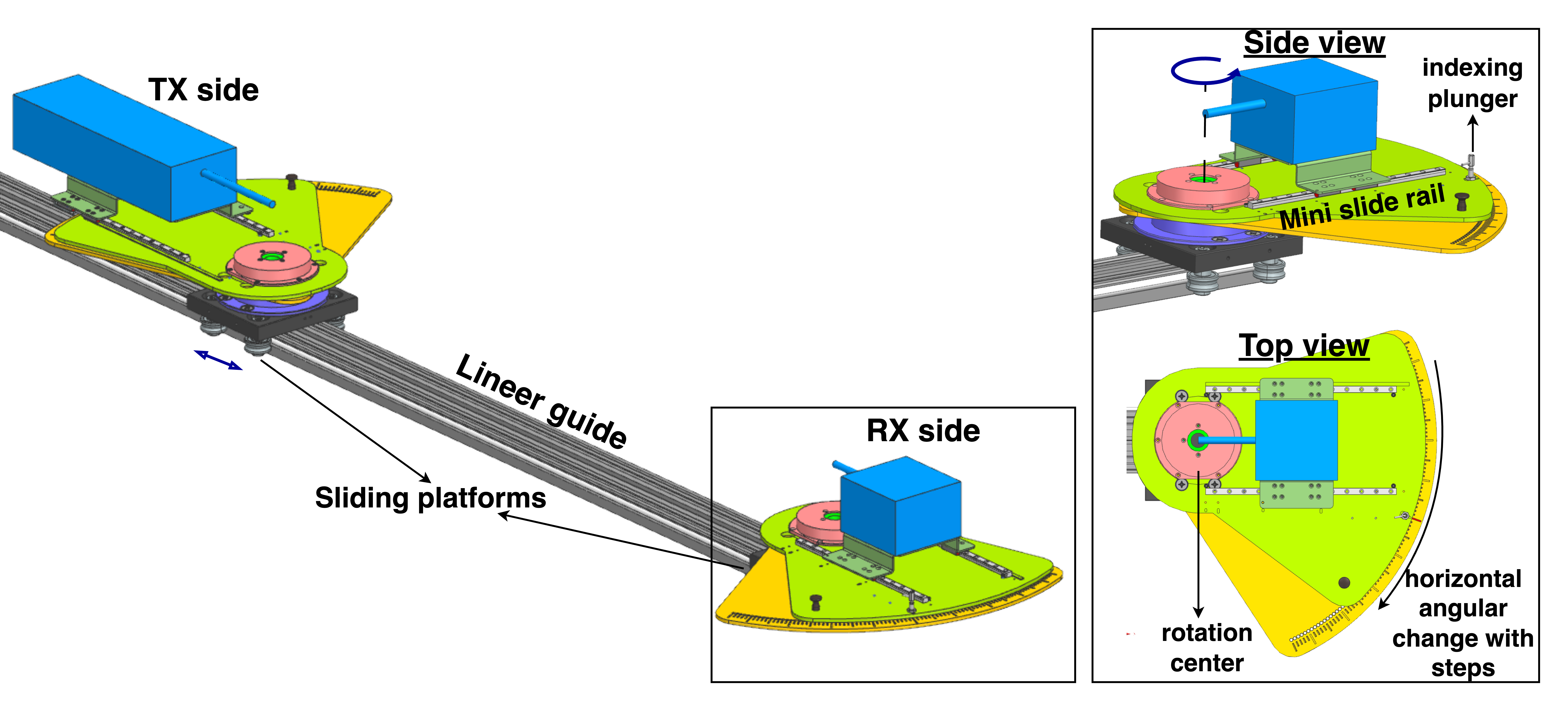}
   \caption{A three-dimensional (3D) model depicting the mechanical components of the measurement setup, including a rotating mechanism with 1-degree precision in the horizontal direction and an adjustable horizontal distance between the transmitter and receiver.}
   \label{fig:autocad}
   \vspace{-10pt}
\end{figure*} 

\subsection{Measurement Setup}
\label{sec:measurement_setup}
One of the most critical factors to be considered in misalignment experiments is ensuring the repeatability of certain scenarios and settings of the measurements. 
Especially when the focus is on \ac{thz} frequencies various factors must be considered where even the slightest change can significantly impact the results. Foremost, maintaining the accuracy of the $0^\circ$ position, serving as the reference for misalignment assessment, is crucial. 
Periodic verification of the alignment at other angles is equally important. Thus, a rotation platform has been engineered to facilitate precise adjustments of horizontal angles, offering seamless transitions at one-degree intervals.
% The antenna must be in the centre of the rotation platform which allows it to keep the distance fixed even when the receiver is rotated. 
To maintain a consistent distance during receiver rotation, it is crucial to place the antenna precisely at the center of the rotation platform.
% To ensure reproducibility in measurements, the movable rotation platform should be arranged so that the calibration is renewed at regular intervals without removing the extenders. 
To guarantee the consistency of measurements, the mobile rotation platform should be configured in a manner that allows for periodic recalibration without the need to disassemble the extenders.
Such a process that starts from designing the measurement setup that serves a particular purpose and reaches the results verified by repeated measurements obliges carrying out a multi-disciplinary effort. Misalignment is controlled by the angular rotation of the receiver. 
% Thanks to the mini slide rail, when the calibration is needed the extenders are brought closer to each other without being removed from the setup, while the indexing plunger prevents unwanted angle changes. 
The mini slide rail facilitates calibration by bringing the extenders closer without necessitating their removal from the setup, with the indexing plunger effectively preventing undesired angle changes.
This measurement setup not only enables the assessment of misalignment but also allows for distance-dependent measurements, given the mobility of the rotation platforms with rails.
% With this measurement setup, besides misalignment, distance-dependent measurement can also be taken since the rotation platforms are movable. 
% Thirdly, it is crucial to ensure effective management of cables to avoid signal deterioration and interference. It is recommended to utilize high-quality cables that have suitable shielding to minimize signal losses and maintain a stable transmission. For this study, Minicircuit brand cables with low loss characteristics are preferred, particularly when operating at high frequencies.
% Also, effective cable management is imperative to prevent signal degradation and interference. Utilizing high-quality cables with appropriate shielding is strongly recommended especially when operating at high frequencies to minimize signal losses and ensure stable transmission. In this study, Minicircuit branded cables, known for their low-loss characteristics, are preferred.
% It is of utmost importance to perform periodic calibration of the measurement equipment too, which includes antennas and extenders, to uphold precision and compensate for any alterations or fluctuations in the equipment's performance. This process involves comparing measured signals with well-established reference standards.
% Furthermore, it is vital to manage control over the experimental environment to reduce the impact of external factors such as temperature, humidity, and electromagnetic interference on the measurements. 
Also, effective cable management is essential to maintain signal integrity and minimize interference. It's essential to use high-quality cables with proper shielding, particularly when operating at high frequencies, to mitigate signal losses and ensure reliable transmission. For this study, Minicircuit branded cables are preferred for their low-loss characteristics.
Likewise, regular calibration of measurement equipment, including antennas and extenders, is paramount to maintain precision and compensate for any variations or fluctuations in performance. This involves comparing measured signals with established reference standards.
Moreover, meticulous control over the experimental environment is vital to minimize the influence of external factors like temperature, humidity, and electromagnetic interference on the measurements.% Implementing shielding measures to safeguard the measurements from external electromagnetic radiation sources is crucial in ensuring precise and accurate measurements. 
By adhering to these design guidelines, a robust measurement setup can be created that actively minimizes the impact of external factors, resulting in reproducible accurate and reliable measurements.

% Thirdly, proper cable management is essential to prevent signal degradation and interference. Using high-quality cables with appropriate shielding to minimize losses and ensure stable signal transmission. In this study, low loss cables of the Minicircuit brand are preferred at high frequencies. Fourtly, regular calibration of the measurement equipment, including antennas and receivers, is crucial for maintaining accuracy. This involves comparing the measured signals against known reference standards or calibrated sources. Periodic recalibration will help account for any changes or drift in the equipment's performance. Fifthly, to minimize external factors that could affect the measurements, it's important to control the environment in which the experiments are conducted. This includes factors such as temperature, humidity, and electromagnetic interference. Shielding the setup from external sources of electromagnetic radiation can help ensure accurate measurements.

%The \ac{thz} measurement setup which allows for evaluating the effect of the misalignment designed by considering all these criteria is depicted in \FGR{fig:autocad}. 
\subsubsection{Description of Measurement Setup}
\begin{figure}[!t]
    \centering    \includegraphics[width=\linewidth]{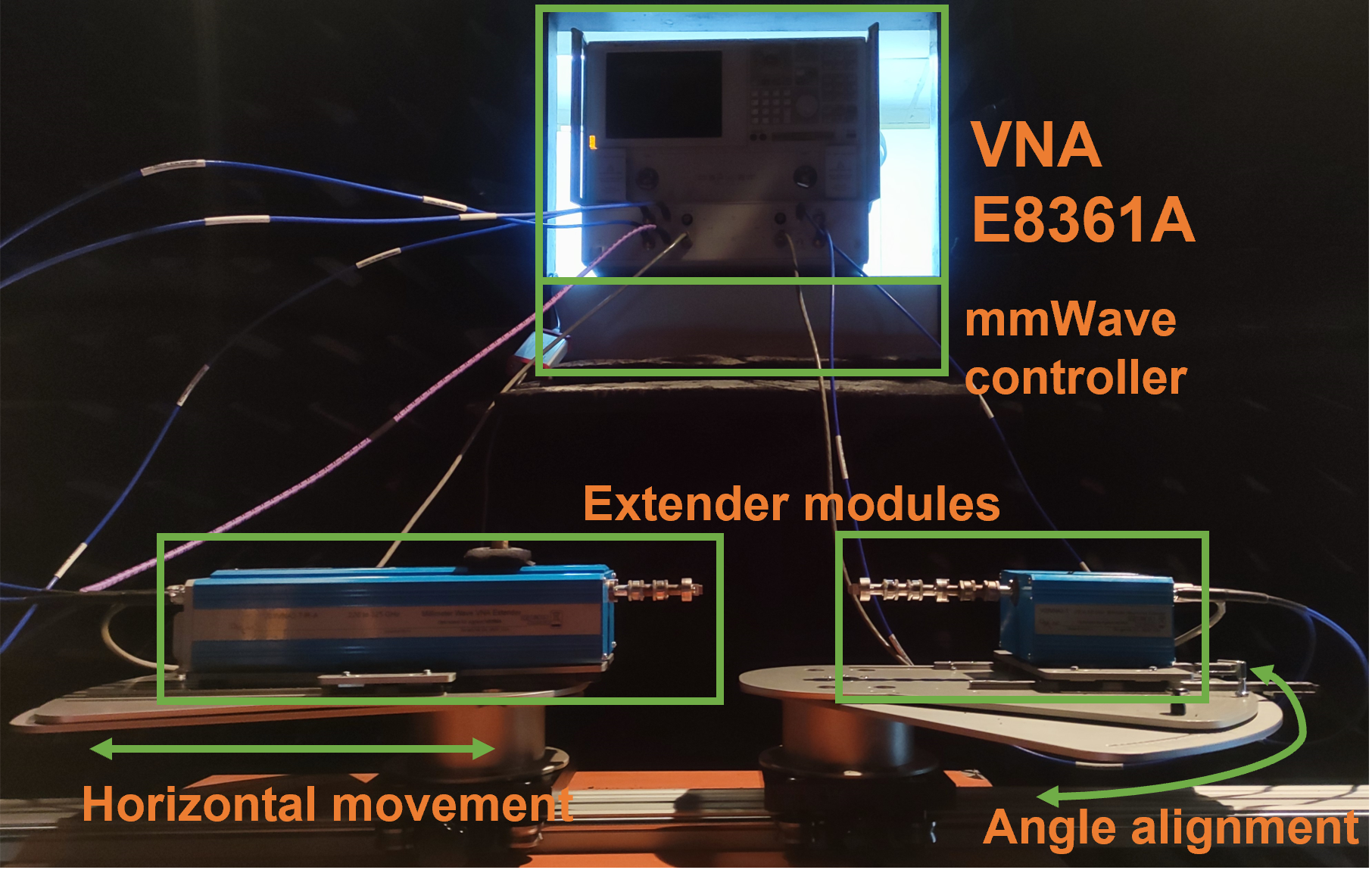}
    \caption{Measurement setup.}
    \label{fig:measurement_setup}
    \vspace{-10pt}
\end{figure}
Our THz experimental setup is shown in \FGR{fig:measurement_setup} which is constructed in the MİLTAL at the TÜBİTAK. The measurement setup consists of four main hardware parts and mechanical parts. Hardware parts of the system consist of Agilent vector network analyzer (VNA) E8361A, Oleson Microwave Labs (OML) V03VNA2-T and V03VNA2–T/R–A millimetre wave extender modules and N5260A extender controller. 
\begin{figure}[!t]
    \centering    \includegraphics[width=\linewidth]{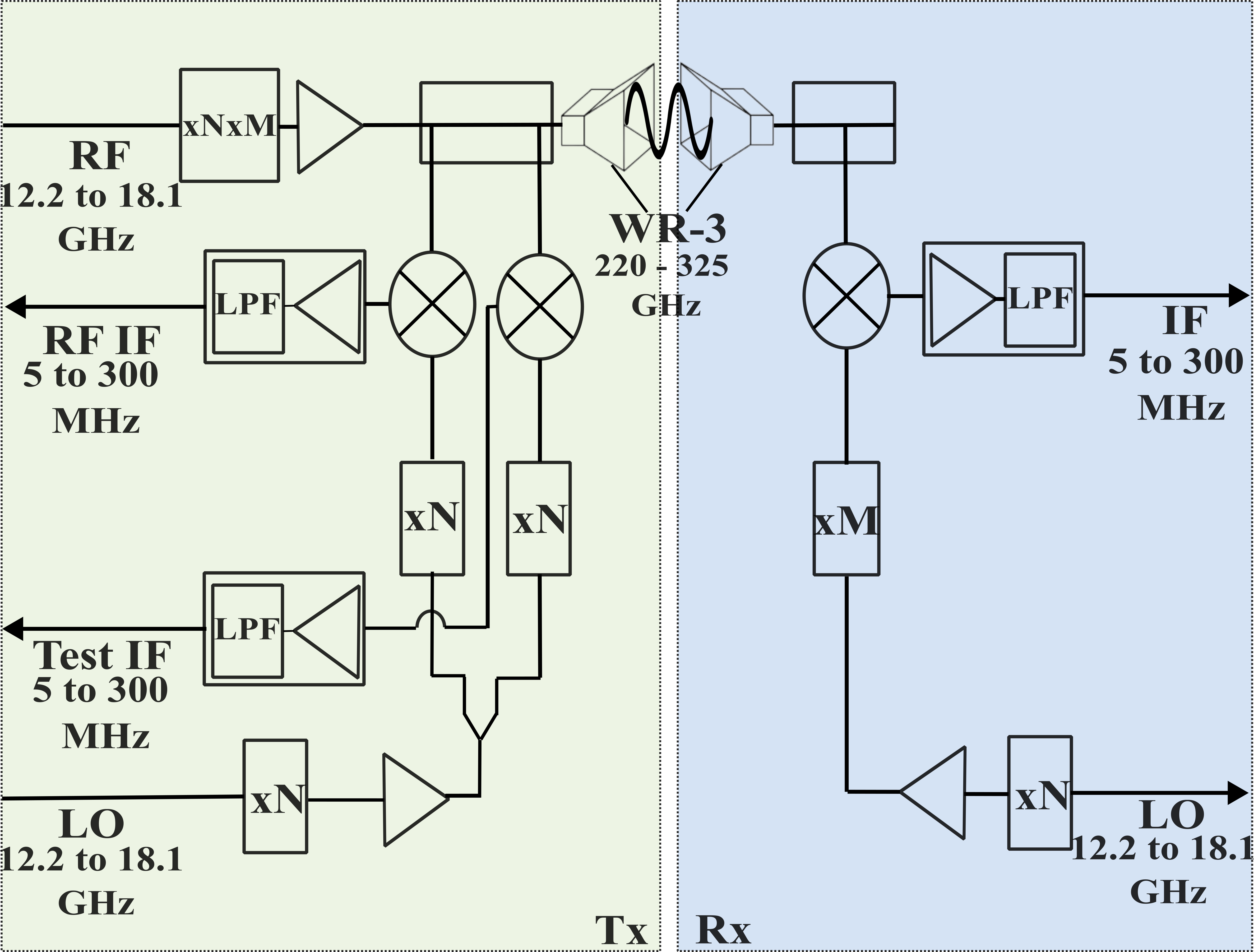}
    \caption{Block diagram of transmitter and receiver extenders.}
    \vspace{5pt}
    \label{fig:)}
    
\end{figure}

% Due to VNA's frequency limit of 67 GHz, we have attached the extender modules, which work between 220 GHz and 325 GHz, to the VNA for analyzing the misalignment effect at the \ac{thz} frequency band. 
% Due to VNA's operational frequency limit of 67 GHz, in order to give the ability to the analyzing misalignment effect on the THz frequency range, the extender modules are coupled with the VNA which allows it the analyzing signals between 220 GHz and 325 GHz.
To enable the analysis of misalignment effects on the THz frequency range, extender modules are coupled with the VNA which is limited to an operational frequency of 67 GHz. 
% The extender modules extend the frequency range of the VNA, allowing it to analyze signals between 220 GHz and 325 GHz.
% We coupled the extender modules, which work between 220 GHz and 325 GHz, to the VNA to analyze the misalignment impact at the \ac{thz} frequency band due to the VNA's 67 GHz frequency limit.
% V03VNA2–T/R–A multiplies the input RF signal between 12.2 GHz and 18.1 GHz with 18 in order to extend the 220 GHz to 325 GHz frequency range. Before transmitting the signal, test intermediate frequency (IF) and reference IF signals for VNA are obtained from downconversion mixers. After that received signal passing through the channel is downconverted at the V03VNA2–T, and the resulting test IF signal is fed back to the VNA. 
The V03VNA2–T/R–A drives up the RF signal within the 12.2 GHz and 18.1 GHz range by a factor of 18 and expands the frequency of the transmission signals allowing VNA to analyze signals between 220 GHz and 325 GHz.
Before transmission, the VNA acquires test intermediate frequency (IF) and reference IF signals through downconversion mixers. 
Following signal reception after passing through the channel, it undergoes downconversion at V03VNA2-T which results in a test IF signal that is fed back to the VNA for further analysis. 
% at the V03VNA2-T following the reception, 
% resulting in a test IF signal that is fed back to the VNA for further analysis.
The extender modules have been driven by an extended band WR-10 multiplier chain with a waveguide output interface. The waveguide output power of the V03VNA2-T/R-A is around -23 dBm. Also, the magnitude and phase stability of the extenders are $\pm$0.4dB and $\pm$8$^\circ$, respectively. Also, a half-power beam width horn antenna with 25 dBi gain \cite{RFechoAntenna} is used in measurements. Because of the narrow beamwidth at high frequencies, the alignment between the transmitter and receiver needs to be very precise. So, the extender modules have been installed in a mechanical system as mentioned in Section \ref{sec:measurement_setup} where we can precisely change the distance and angles between the modules. 

% In this study, the operating frequency range is set to 240 GHz to 300 GHz because the magnitude and phase stability of the extender modules has better performance in this range. Before measuring the s-parameters, we calibrated the extender modules by connecting the transmitter and the receiver module's waveguide ports. 
% Following calibration, 4096 points with an IF bandwidth of 100 Hz are used to measure the 60 GHz band.
% After the calibration process, the 60 GHz band is measured with 4096 points averaging 100 Hz IF bandwidth.

\begin{table}[]
\caption{Measurement Parameters}
\label{tab:measurement_parameters}
\centering
\renewcommand{\arraystretch}{1.2}
\setlength{\tabcolsep}{4pt}
\begin{tabular}{ll}
\hline
Description                  & Value                                                                                                                       \\ \hline \hline
Operating frequency          & 240GHz - 300GHz                                                                                                             \\
Bandwidth                    & 60GHz                                                                                                                       \\
Measurement frequency points & 4096                                                                                                                        \\
IF bandwidth                 & 100Hz                                                                                                                       \\
Spectral resolution          & 14.648MHz                                                                                                                   \\
Antenna misalignment         & $0^\circ:1^\circ:15^\circ$ and $15^\circ:5^\circ:30^\circ$  \\
Distance (cm)                & 20:10:100cm                                                                                  \\ \hline
\end{tabular}
\vspace{-15pt}
\end{table}

\begin{figure}[t!]
    \centering
    \includegraphics[width=\linewidth]{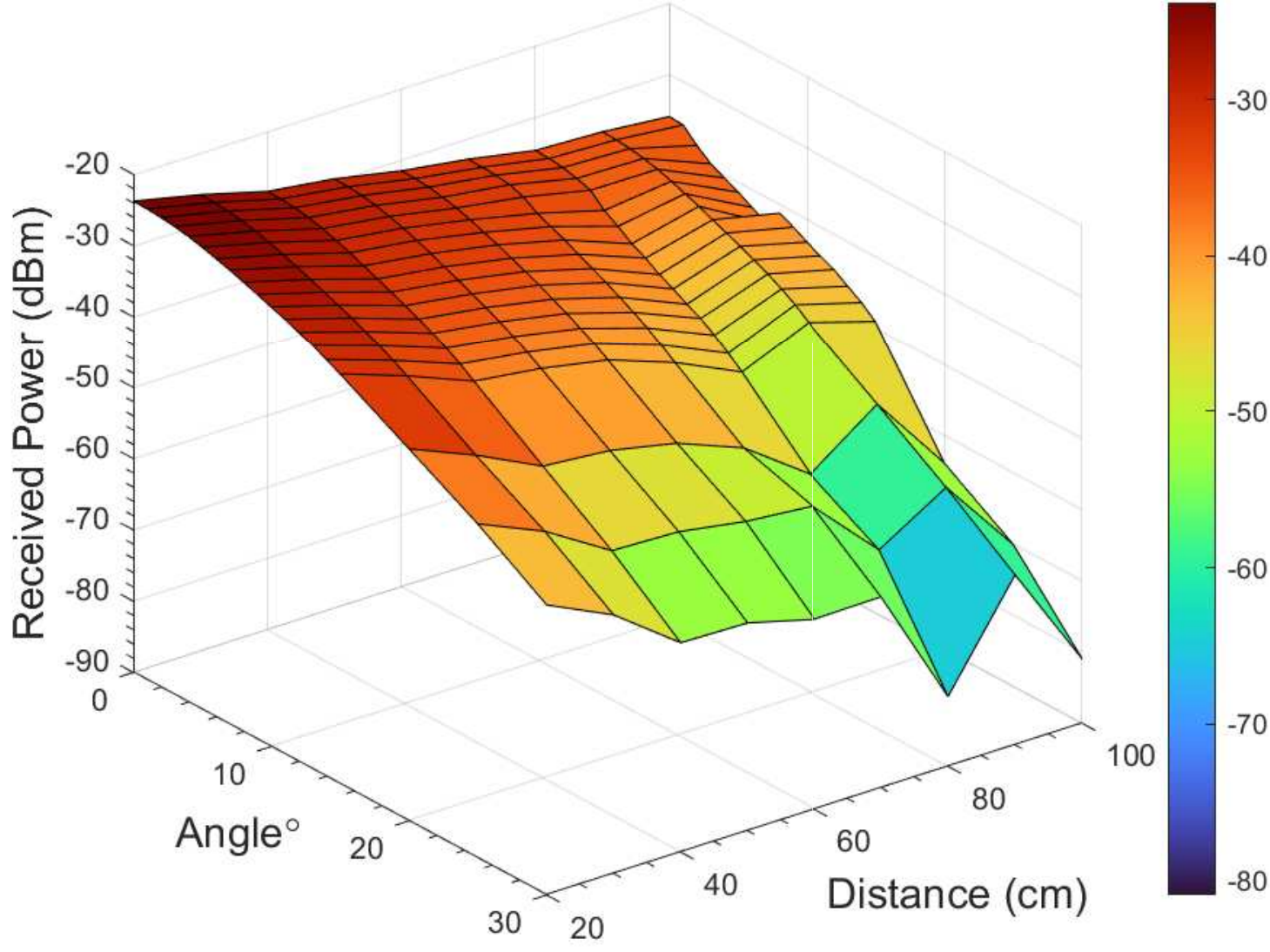}
    \caption{Maximum values of the channel impulse response for all distances and antenna misalignment degrees.}
    \label{fig:cir}
    \vspace{-10pt}
\end{figure}
\subsubsection{Measurement Methodology}
In this study, the operating frequency range is set to 240 GHz to 300 GHz because of the magnitude and phase stability of the extender modules in this range. 
To ensure accuracy, the spectral resolution of each measurement is set to be 14.648 MHz, which corresponds to 4096 frequency points with an IF bandwidth of 100 Hz. 
% To ensure measurements are reliable, electronic devices including cables calibrated together by connecting the transmitter and the receiver module's waveguide ports end-to-end to retrieve unwanted effects of the electronics before the measurements. 
To ensure accurate measurements, it is necessary to calibrate electronic devices and cables together. Prior to the measurements, calibration was done by connecting the waveguide ports of the transmitter and receiver modules end-to-end to retrieve any unwanted effects caused by the electronics. 
% Before the measurements, electronic devices including cables calibrated together by connecting the transmitter and the receiver module's waveguide ports end-to-end to retrieve unwanted effects of the electronics. 
So as to comprehensively investigate the impact of antenna misalignment on received power in the \ac{thz} wireless channel, a series of measurements were conducted using a sliding rail and rotation platform.
These platforms enabled horizontal angle adjustments with a precision of $1^{\circ}$ at each distance setting, allowing for a comprehensive investigation of the influence of antenna misalignment with distance.

% To investigate exhaustively the impact of antenna misalignment with distance on received power in the \ac{thz} wireless channel, several measurements from different settings are obtained with the help of sliding rail and rotation platform which allows horizontal angle adjustment with the precision of $1^{\circ}$ at each distance setting. 

In order to facilitate analysis and improve the reliability of data by minimizing the number of unknown variables, this study specifically evaluated changes in only the horizontal angle.
By narrowing the scope to changes in the horizontal angle, the research yielded more precise and trustworthy findings.
% By limiting the focus to horizontal angle changes, the study was able to produce more accurate and dependable findings. 
The measurements were obtained as $S_{21}$ parameters and stored as complex numbers in Agilent VNA E8361A. The parameters of the measurements are presented in detail in Table \ref{tab:measurement_parameters}.

\begin{figure}[t!]
    \centering
    \includegraphics[width=\linewidth]{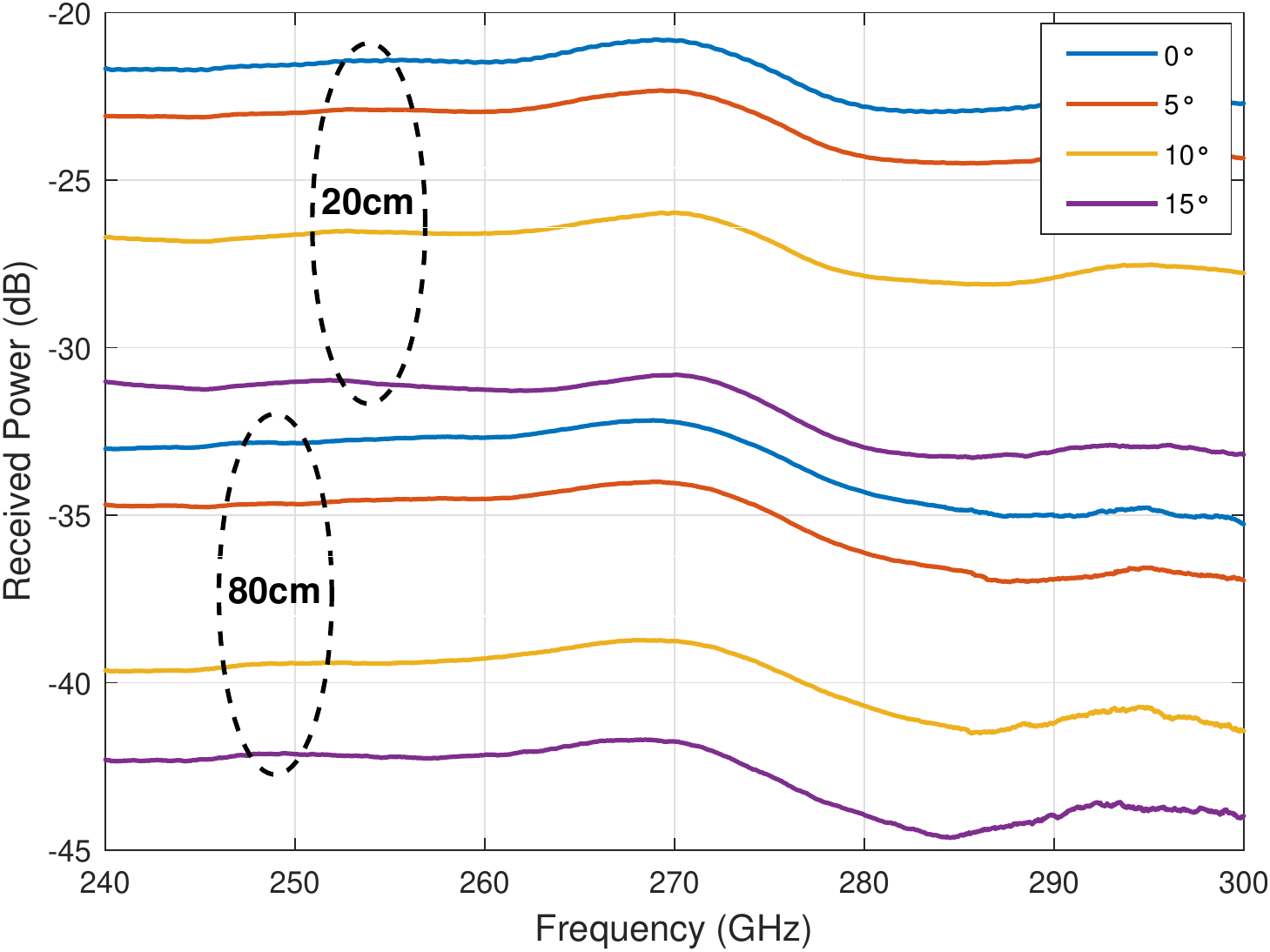}
    \caption{Channel frequency responses with $0^\circ, 5^\circ, 10^\circ$ and $15^\circ$ antenna misalignment at 20cm and 80cm distances.}
    \label{fig:cfr}
\end{figure}

\begin{figure}[t!]
    \centering
   \includegraphics[width=\linewidth]{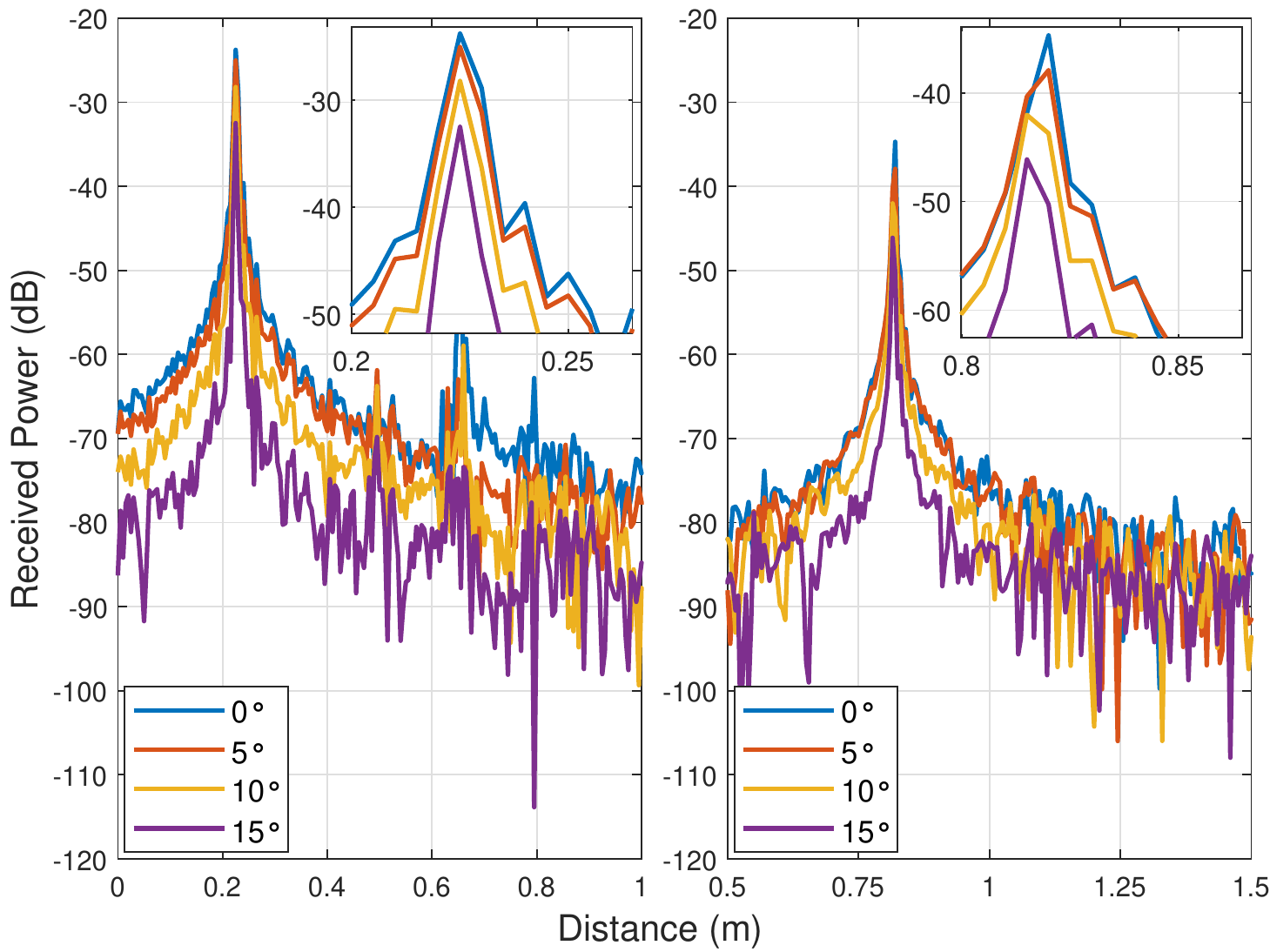}
    \caption{Channel impulse responses with $0^\circ, 5^\circ, 10^\circ$ and $15^\circ$ antenna misalignment at 20 cm and 80 cm distances.}
    \label{fig:cir_seperated}
    \vspace{-10pt}
\end{figure}

\begin{table}[]
\caption{Received signal peak powers at all distances and antenna misalignments}
\label{tab:cir_values}
\centering
\renewcommand{\arraystretch}{1.2}
\setlength{\tabcolsep}{4pt}
\begin{tabular}{cccccccc}
\hline
\multicolumn{1}{l}{} & 0${^\circ}$ & 5$^{\circ}$ & 10$^{\circ}$ & 15$^{\circ}$ & 20$^{\circ}$ & 25$^{\circ}$ & 30$^{\circ}$ \\ \hline \hline
20 cm                & -23.77      & -25.02      & -28.19       & -32.48       & -37.79       & -43.13       & -49.29       \\
30 cm                & -25.86      & -27.35      & -30.95       & -35.78       & -41.71       & -47.17       & -53.70       \\
40 cm                & -28.42      & -30.07      & -34.02       & -39.52       & -46.24       & -52.90       & -56.92       \\
50 cm                & -29.66      & -31.38      & -35.40       & -40.74       & -47.07       & -53.25       & -60.85       \\
60 cm                & -31.55      & -33.21      & -37.16       & -42.56       & -49.05       & -54.82       & -62.07       \\
70 cm                & -32.92      & -35.32      & -39.82       & -45.93       & -52.73       & -55.68       & -63.20       \\
80 cm                & -34.71      & -37.92      & -42.03       & -46.13       & -50.92       & -56.15       & -60.51       \\
90 cm                & -35.11      & -36.84      & -40.76       & -46.27       & -52.59       & -59.08       & -66.16       \\
100 cm               & -35.98      & -39.92      & -43.17       & -49.14       & -52.67       & -57.57       & -63.02       \\ \hline
\end{tabular}
\vspace{-10pt}
\end{table}

\section{Measurement Results}\label{measurement_results}
In this section, the joint impact of the antenna misalignment and the distance dependent path loss is presented by illustrating the channel frequency response and impulse response of the measurements. 
The channel frequency responses for 0,3,5,10 and 15-degree antenna misalignment at 20 cm and 80 cm are shown in Fig. \ref{fig:cfr}.
% When only distance is considered without any misalignment the received power changes around 12 dB when the separation is increased from 20 cm to 80 cm. 
If only the distance between the transmitter and receiver is taken into account, the received power experiences a change of approximately 12 dB when the separation is increased from 20 cm to 80 cm.
The 15-degree horizontal angle change from the reference point at 20 cm causes the 8 dB reduction of the received power.

For example, the received power difference is around 12 dB without antenna misalignment for 20 cm and 80 cm. In case of antenna misalignment, loss of the received power can reach up to 8 dB. In addition, \FGR{fig:cir_seperated} shows that time domain analysis by taking inverse Fourier Transform of the measurement data. In this figure, it is seen that there is a decrease in received power due to antenna misalignment. So, antenna misalignment considerably impacts received signal power because \ac{thz} antennas have a narrow beam width.
Furthermore, \FGR{fig:cir} illustrates the channel impulse response with the combined distance and antenna misalignment measurements. It is plotted using the max value of the channel impulse response for every distance and angle measurement pair. In addition, the direct numerical value equivalents of these measurement pairs are given in the \TAB{tab:cir_values}. 
% Considering that, by increasing the distance between the transmitter and the receiver from 20 cm to 80 cm, the loss in received power is almost equal to the loss due to 10-degree antenna misalignment at a fixed 80 cm distance, and this loss is approximately 9 dB.
Given that increasing the separation between the transmitter and the receiver from 20 cm to 80 cm results in a received power loss nearly equivalent to the loss incurred by a 10-degree antenna misalignment at a distance of 80 cm, the estimated loss in both scenarios is approximately 9 dB.
%When we compared the distance and misalignment effect to the received power to understand the importance of the antenna misalignment, the received power decreases 9dB from 20cm to 80cm, corresponding to the $10^\circ$ misalignment at both 20cm and the 80cm distance. Despite the square increase depending on the path loss distance, the received power at 60cm distance difference and only $10^\circ$ misalignment received power without changing from the distance are approximately at the same level.  

% In order to better understand the importance of the effect of antenna misalignment on received power, we can analyze the numerical data in \TAB{tab:cir_values}. 
% For example, by increasing the distance between the transmitter and the receiver from 20 cm to 80 cm, the loss in received power is almost equal to the loss due to 10-degree antenna misalignment at a fixed 80 cm distance, and this loss is approximately 9 dB. 
Antenna misalignment is an indispensable concern for \ac{thz} communication systems in \ac{uav}, as it directly impacts the received power. 
The distance between the transmitter and receiver is the primary factor in diminishing received power, with antenna misalignment exacerbating this power loss and causing a further decrease in received power.
% The distance between the transmitter and the receiver plays a crucial role in path loss, which inevitably results in a decline in the received power. In addition, when the transmitter and receiver antennas are misaligned, the issue of power loss is exacerbated, causing a further decrease in the received power. 
In this context, the development of fast and robust beamforming and beam tracking algorithms is imperative for UAVs equipped with THz communication systems.
Compensating for misalignment is crucial for various reasons, ensuring a consistent level of received power and reliable communication in UAVs operating at THz frequencies.
% The compensation for misalignment for several reasons is imperative to maintain the desired level of received power, thus ensuring reliable communication in UAVs operating at THz frequencies.
\section{Conclusion and Future Directions}\label{conclusion}
THz communication systems hold promise as a solution for addressing the high data rate demands and increasing number of wireless devices, and UAVs have been featured to enable ubiquitous access to the sublime potential of these frequencies. 
This study initiates a discussion on antenna misalignment in UAV-assisted THz communication, which will be one of the most critical challenges in the practical implementation step. Experiments were conducted with a fine-tuned setup to explore the impact of horizontal misalignment and distance on received power between 240 GHz and 300 GHz, revealing a significant impact on SNR even minor deviations in alignment. To guide future research, essential factors for misalignment experiments are outlined. 
The next step involves examining both horizontal and vertical misalignment factors for their joint effects on received power.
Furthermore, with advancements in transceiver technology and the feasibility of UAV-mounted THz campaigns, real-world experiments should be conducted.

\section*{Acknowledgment}
This work has received funding from the AIMS5.0 project. AIMS5.0 has been accepted for funding within the Key Digital Technologies Joint Undertaking (KDT JU), a public-private partnership in collaboration with the HORIZON Framework Programme and the national Authorities under grant agreement number 101112089.

% % % % % % % % % % % % % % % % % % % % % % % % % % % %
\balance
\bibliographystyle{IEEEtran}
\bibliography{main.bib}

\begin{thebibliography}{10}
\providecommand{\url}[1]{#1}
\csname url@samestyle\endcsname
\providecommand{\newblock}{\relax}
\providecommand{\bibinfo}[2]{#2}
\providecommand{\BIBentrySTDinterwordspacing}{\spaceskip=0pt\relax}
\providecommand{\BIBentryALTinterwordstretchfactor}{4}
\providecommand{\BIBentryALTinterwordspacing}{\spaceskip=\fontdimen2\font plus
\BIBentryALTinterwordstretchfactor\fontdimen3\font minus \fontdimen4\font\relax}
\providecommand{\BIBforeignlanguage}[2]{{%
\expandafter\ifx\csname l@#1\endcsname\relax
\typeout{** WARNING: IEEEtran.bst: No hyphenation pattern has been}%
\typeout{** loaded for the language `#1'. Using the pattern for}%
\typeout{** the default language instead.}%
\else
\language=\csname l@#1\endcsname
\fi
#2}}
\providecommand{\BIBdecl}{\relax}
\BIBdecl

\bibitem{7417609}
M.~Mozaffari, W.~Saad, M.~Bennis, and M.~Debbah, ``Drone small cells in the clouds: Design, deployment and performance analysis,'' in \emph{2015 IEEE Global Communications Conference (GLOBECOM)}, 2015, pp. 1--6.

\bibitem{sobouti2023managing}
M.~J. Sobouti, A.~H. Mohajerzadeh, S.~A.~H. Seno, and H.~Yanikomeroglu, ``Managing sets of flying base stations using energy efficient {3D} trajectory planning in cellular networks,'' \emph{IEEE Sensors Journal}, 2023.

\bibitem{giordani2020toward}
M.~Giordani, M.~Polese, M.~Mezzavilla, S.~Rangan, and M.~Zorzi, ``Toward {6G} networks: Use cases and technologies,'' \emph{IEEE Communications Magazine}, vol.~58, no.~3, pp. 55--61, 2020.

\bibitem{boulogeorgos2018terahertz}
A.-A.~A. Boulogeorgos, A.~Alexiou, T.~Merkle, C.~Schubert, R.~Elschner, A.~Katsiotis, P.~Stavrianos, D.~Kritharidis, P.-K. Chartsias, J.~Kokkoniemi \emph{et~al.}, ``Terahertz technologies to deliver optical network quality of experience in wireless systems beyond {5G},'' \emph{IEEE Communications Magazine}, vol.~56, no.~6, pp. 144--151, 2018.

\bibitem{9269928}
K.~Tekbiyik, A.~R. Ekti, G.~K. Kurt, A.~Gorcin, and H.~Yanikomeroglu, ``A holistic investigation of terahertz propagation and channel modeling toward vertical heterogeneous networks,'' \emph{IEEE Communications Magazine}, vol.~58, no.~11, pp. 14--20, 2020.

\bibitem{jornet2011channel}
J.~M. Jornet and I.~F. Akyildiz, ``Channel modeling and capacity analysis for electromagnetic wireless nanonetworks in the terahertz band,'' \emph{IEEE Transactions on Wireless Communications}, vol.~10, no.~10, pp. 3211--3221, 2011.

\bibitem{azari2022thz}
M.~M. Azari, S.~Solanki, S.~Chatzinotas, and M.~Bennis, ``{THz}-empowered {UAVs} in {6G}: Opportunities, challenges, and trade-offs,'' \emph{{IEEE Communications Magazine}}, vol.~60, no.~5, pp. 24--30, 2022.

\bibitem{azari2022evolution}
M.~M. Azari, S.~Solanki, S.~Chatzinotas, O.~Kodheli, H.~Sallouha, A.~Colpaert, J.~F.~M. Montoya, S.~Pollin, A.~Haqiqatnejad, A.~Mostaani \emph{et~al.}, ``Evolution of non-terrestrial networks from {5G} to {6G}: A survey,'' \emph{IEEE Communications Surveys \& Tutorials}, 2022.

\bibitem{priebe2012impact}
S.~Priebe, M.~Jacob, and T.~K{\"u}rner, ``The impact of antenna directivities on {THz} indoor channel characteristics,'' in \emph{2012 6th European Conference on Antennas and Propagation (EUCAP)}.\hskip 1em plus 0.5em minus 0.4em\relax IEEE, 2012, pp. 478--482.

\bibitem{priebe2012affection}
------, ``Affection of {THz} indoor communication links by antenna misalignment,'' in \emph{2012 6th European Conference on Antennas and Propagation (EUCAP)}.\hskip 1em plus 0.5em minus 0.4em\relax IEEE, 2012, pp. 483--487.

\bibitem{papasotiriou2020performance}
E.~N. Papasotiriou, A.-A.~A. Boulogeorgos, and A.~Alexiou, ``Performance analysis of {THz} wireless systems in the presence of antenna misalignment and phase noise,'' \emph{IEEE Communications Letters}, vol.~24, no.~6, pp. 1211--1215, 2020.

\bibitem{badarneh2022performance}
O.~S. Badarneh, ``Performance analysis of terahertz communications in random fog conditions with misalignment,'' \emph{IEEE Wireless Communications Letters}, vol.~11, no.~5, pp. 962--966, 2022.

\bibitem{ekti2017statistical}
A.~R. Ekti, A.~Boyaci, A.~Alparslan, {\.I}.~{\"U}nal, S.~Yarkan, A.~G{\"o}r{\c{c}}in, H.~Arslan, and M.~Uysal, ``Statistical modeling of propagation channels for terahertz band,'' in \emph{2017 IEEE Conference on Standards for Communications and Networking (CSCN)}.\hskip 1em plus 0.5em minus 0.4em\relax IEEE, 2017, pp. 275--280.

\bibitem{sheikh2021horn}
F.~Sheikh, Y.~Zantah, M.~Al-Hasan, I.~Mabrouk, N.~Zarifeh, and T.~Kaiser, ``Horn antenna misalignments at 100, 300, 400, and 500 ghz in close proximity communications,'' in \emph{2021 IEEE International Symposium on Antennas and Propagation and USNC-URSI Radio Science Meeting (APS/URSI)}.\hskip 1em plus 0.5em minus 0.4em\relax IEEE, 2021, pp. 449--450.

\bibitem{9852427}
E.~Karakoca, G.~Karabulut~Kurt, and A.~Görçi̇n, ``Hierarchical {D}irichlet process based {G}amma mixture modeling for terahertz band wireless communication channels,'' \emph{IEEE Access}, vol.~10, pp. 84\,635--84\,647, 2022.

\bibitem{10437716}
E.~Karakoca, H.~Nayir, G.~K. Kurt, and A.~Görçin, ``Measurement-based modeling of short range terahertz channels and their capacity analysis,'' in \emph{GLOBECOM 2023 - 2023 IEEE Global Communications Conference}, 2023, pp. 1471--1476.

\bibitem{guan2019effects}
Z.~Guan and T.~Kulkarni, ``On the effects of mobility uncertainties on wireless communications between flying drones in the mmwave/thz bands,'' in \emph{IEEE INFOCOM 2019-IEEE Conference on Computer Communications Workshops (INFOCOM WKSHPS)}.\hskip 1em plus 0.5em minus 0.4em\relax IEEE, 2019, pp. 768--773.

\bibitem{yarkan2008identification}
S.~Yarkan and H.~Arslan, ``Identification of los in time-varying, frequency selective radio channels,'' \emph{EURASIP Journal on wireless communications and networking}, vol. 2008, pp. 1--14, 2008.

\bibitem{zang2010bayesian}
H.~Zang, F.~Baccelli, and J.~Bolot, ``Bayesian inference for localization in cellular networks,'' in \emph{2010 Proceedings IEEE INFOCOM}.\hskip 1em plus 0.5em minus 0.4em\relax IEEE, 2010, pp. 1--9.

\bibitem{RFechoAntenna}
RFecho, ``{25 dBi Gain 220GHz-325GHz WR-03 Waveguide Millimeter SGH Antenna},'' {\scriptsize{\url{https://www.rfecho.com/product/25-dbi-gain-220-ghz-to-325-ghz-wr-03-waveguide-millimeter-sgh-antenna}}}, [Online; accessed: October 19, 2023].

\end{thebibliography}
% % % % % % % % % % % % % % % % % % % % % % % % % % % %

\end{document}